# The water abundance in Jupiter's equatorial zone

Cheng Li[1], Andrew Ingersoll[1], Scott Bolton[2], Steven Levin[3], Michael Janssen[3], Sushil Atreya[4], Jonathan Lunine[5], Paul Steffes[6], Shannon Brown[3], Tristan Guillot[7], Michael Allison[8], John Arballo[3], Amadeo Bellotti[9], Virgil Adumitroaie[3], Samuel Gulkis[3], Amoree Hodges[6], Liming Li[10], Sidharth Misra[3], Glenn Orton[3], Fabiano Oyafuso[3], Daniel Santos-Costa[2], Hunter Waite[2], Zhimeng Zhang[1]

[1]California Institute of Technology, Pasadena CA, 91125

[2]Southwest Research Institute, San Antonio TX, 78228

[3]Jet Propulsion Laboratory, California Institute of Technology, Pasadena CA, 91108

[4]University of Michigan, Ann Arbor MI, 48109

[5]Cornell University, Ithaca NY, 14850

[6]Georgia Institute of Technology, Atlanta GA, 30332

[7]Universite Cote d'Azur, OCA, Lagrange CNRS, 06304 Nice, France

[8]Goddard Institute for Space Studies, New York NY, 10025

[9]Lockheed Martin, Grand Prairie TX, 75208

[10]University of Houston, Houston TX, 77004

Corresponding author: Cheng Li (cli@gps.caltech.edu)

**Oxygen is the most common element after hydrogen and helium in Jupiter's atmosphere, and may have been the primary condensable (as water ice) in the protoplanetary disk. Prior to the Juno mission, in situ measurements of Jupiter's water abundance were obtained from the Galileo Probe, which dropped into a meteorologically anomalous site. The findings of the Galileo Probe were inconclusive because the concentration of water was still increasing when the probe died. Here, we initially report on the water abundance in the equatorial region, from 0 to 4 degrees north latitude, based on 1.25 to 22 GHz data from Juno Microwave radiometer probing approximately 0.7 to 30 bars pressure. Because Juno discovered the deep atmosphere to be surprisingly variable as a function of latitude, it remains to confirm whether the equatorial abundance represents Jupiter's global water abundance. The water abundance at the equatorial region is inferred to be $2.5^{+2.2}_{-1.6} \times 10^{3}$ ppm, or $2.7^{+2.4}_{-1.7}$ times the protosolar oxygen elemental ratio to H (1σ uncertainties). If reflective of the global water abundance, the result suggests that the planetesimals formed Jupiter are unlikely to be water-rich clathrate hydrates.**

From thermodynamic calculations[1], three types of cloud layers in the Jovian atmosphere are thought to exist: an ammonia ice cloud, an ammonium hydrosulfide ice cloud[2,3], and a water ice and droplet cloud, formed approximately at 0.7 bars, 2.2 bars, and 5 bars, respectively, assuming solar abundances. The locations of these clouds may vary due to the local abundance, meteorology and specific model parameters. Condensation and evaporation of water contribute to weather on giant planets because water is the most abundant species apart from hydrogen and helium and the latent heat flux in convective storms is comparable to the solar and internal heat fluxes[4,5]. Consequently, the thermal state of the atmosphere is affected by the amount of water vapor in the atmosphere. Prior to the Juno mission, in situ measurements of Jupiter's atmospheric composition below the clouds were obtained from the Galileo Probe[6], which dropped into a meteorologically anomalous site (6.57° N planetocentric latitude , 4.46° W longitude)[7], known as a 5 $\mu m$ "hot spot" near the boundary between the visibly-bright Equatorial Zone (EZ) and the dark North Equatorial Belt (NEB)[8]. The findings of the Galileo Probe were baffling, for they showed that the levels where ammonia and hydrogen sulfide become uniformly mixed occur much deeper (~10 bars) than what was predicted by an equilibrium thermochemical model. The concentration of water was subsolar and still increasing at 22 bars, where radio contact with the probe was lost, although the concentrations of nitrogen and sulfur stabilized at ~3 times solar at ~10 bars[9,10]. The depletion of water was proposed to be caused by meteorology at the probe location[8,11]. The observed water abundance was assumed not to represent the global mean water abundance on Jupiter, which is an important quantity that distinguishes planetary formation models[12–16] and affects atmospheric thermal structure[17,18]. The Juno mission was in part motivated by the necessity of determining the water abundance at multiple locations across the planet. Here we report on the initial analysis of the equatorial zone (EZ) defined as 0 to +4 degrees latitude, the first of Jupiter's regions analyzed.

We analyze the first eight of Juno's orbits around Jupiter (designated as PJ1, PJ3, PJ4, PJ5, PJ6, PJ7, PJ8 and PJ9), with each perijove probing a cross-section of Jupiter's atmosphere at a different longitude and spanning latitudes from the north pole to the south pole (no MWR data were obtained during PJ2). Dates and longitudes of each perijove are summarized in Supplementary Table 1. The raw data (antenna temperatures) were processed through a deconvolution algorithm (See Methods and Janssen et al. , 2017[19]) that removes synchrotron radiation and microwave cosmic background radiation to recover calibrated atmospheric brightness temperatures whose emission angle dependence is parameterized by three coefficients with spatial resolution constrained by geometry and the antenna beam. The subset of nadir brightness temperatures thus obtained from PJ1 to PJ9 is displayed in Fig. 1. The spatial resolution is highest near the equator (~0.5 degree) and lowest toward the poles because the spacecraft is closer to the planet near the equator (~ 4000 km above 1 bar level) and further away near the poles. Despite an approximate 45-degree separation in longitude between each of the PJ1 to PJ9 orbits, the observed nadir brightness temperatures show extremely good consistency, especially at latitudes near the equator, at mid-latitudes, and at pressure levels larger than 10 bars. Compared to the simulated brightness temperatures of a reference adiabatic atmosphere, the observed nadir brightness temperatures are warmer than expected at all channels everywhere except within a narrow latitudinal range of a few degrees located near the equator.

Our new analysis has extended the latitudinal coverage to the poles, while our previous analysis based on PJ1 data[20] was confined to within 40 degrees of the equator.

Fig. 1 shows that the values of the brightness temperature in the EZ are consistent with an ideal moist adiabat in which the temperature is moist adiabatic and the condensing species are well-mixed up to their condensation level. This is also in line with the lack of lightning observations in this region[21] because an ideal moist adiabatic temperature profile does not have enough convective available potential energy to produce lightning. A free inversion of the vertical distribution of ammonia gas using observed nadir PJ1 data[20,22] produced a nearly uniform distribution of ammonia gas, from ~100 bars to 1 bar within a few degrees of the equator, which further corroborates the observation that the ammonia gas in the EZ is mixed vertically to a greater degree than at other latitudes. The consistency between multiple perijoves confirms that the uniformity of ammonia in the EZ is independent of longitude. In addition, similar enrichment of ammonia gas in the EZ relative to the other latitudes was also found by the Cassini/CIRS retrievals[23] and VLA observations[24] although they were sensitive to only shallow levels and pressures less than a few bars. Thus, we identify the EZ as a unique region appropriate for deriving the water abundance assuming that the temperature profile follows a moist adiabat.

The MWR instrument measures the brightness temperature both at nadir and off-nadir angles. The precision of the angular dependence of the brightness temperature provides important additional information and a tight constraint on atmospheric structure beyond that available from the nadir brightness temperature alone. This is due to the fact that an absolute calibration uncertainty of about 2% does not affect the relative brightness at different angles[19]. We define the limb-darkening parameter, $R(\theta)$, as the fractional reduction of the brightness temperature when viewing off-nadir:

$$R(\theta) = \frac{T_b(\theta) - T_b(0)}{T_b(0)}, \qquad (1)$$

Where $T_b(\theta)$ is the brightness temperature at emission angle $\theta$. Fig. 2 shows the limb-darkening at an emission angle of 45 degrees versus the nadir brightness temperature of all channels. Different colors identify different orbits, from PJ1 to PJ9. Error bars associated with each dot represent one sigma uncertainties due to the deconvolution and the instrument noise. The scattering of dots shows the variability of the atmosphere over longitude and time. It should be noted that three types of orbital attitudes are used. For the MWR and MWR/tilt orbits, the spacecraft is oriented so that the scan circle of the antenna beams always includes Jupiter nadir. For the gravity orbits, the spacecraft spin axis is oriented toward the earth, resulting in larger minimum emission angles as the orbit precesses. The consistency of multiple and different kinds of orbits and the absence of systematic correlation between the nadir brightness temperature and the limb-darkening in 0.6 – 5.2 GHz channels indicate that nadir brightness temperature and limb-darkening are two independent measures of the thermal and compositional properties of the atmosphere. Nadir brightness temperature and limb-darkening have larger correlations in the 10 and 22 GHz channels because the limb-darkening is small in these channels due to narrower weighting functions and its estimation, by the deconvolution process, is influenced by the nadir brightness temperature.

Even though the EZ region is characterized by a greater degree of mixing, MWR data indicate possible departures from an ideal moist adiabat. MWR observations at the 5.2 and 10 GHz channels (Fig. 1) measure slightly colder brightness temperatures (by ~ 5 K) than the ideal moist adiabatic model. The potential change in temperature due to ammonia condensation/vaporization at these levels is only $x\mathrm{NH_3}L\mathrm{NH_3}/c_{P,\mathrm{H_2}} \approx 0.3$ K – where $x\mathrm{NH_3}$ is the ammonia mole fraction, $L\mathrm{NH_3}$ is the latent heat of ammonia, and $c_{P,\mathrm{H_2}}$ is the specific heat of $\mathrm{H_2}$ at constant pressure – and therefore does not account for the observations. To account for the coldness, we allow the possibility that ammonia abundance is enriched by 10 to 15% (parameterized by the factor $f_x$ in equation (3) with respect to the deep abundance near a few bars). Simulation of two-dimensional moist convection in Jovian atmospheres[25] suggests that the slight enrichment of ammonia in the subcloud layer is made possible by evaporated ammonia precipitation. The cold plumes of ammonia-rich precipitation sink into the atmosphere to a few bars, where they encounter stable layers due to an increase in the water abundance. This creates a local maximum in ammonia concentration – referred to as the enriched ammonia layer – that is consistent with the observations. As a result, the model for the ammonia distribution consists of five parameters: 1) the ammonia enrichment factor $f_x$, 2) the pressure of the enriched ammonia layer $p_D$, 3) its thickness $\Delta p$, 4) the deep ammonia abundance $x\mathrm{NH}_3$, and 5) the deep water abundance $x\mathrm{H}_2\mathrm{O}$. Since we do not know the functional form of ammonia profile, $q(p)$, in the enriched ammonia layer, we use an exponential profile that results from balancing the downward diffusive flux and upward advective flux. Its functional form in pressure coordinates is:

$$q(p) = A + B\exp\left(-\frac{p - p_D}{\Delta p}\right), p > p_D, \tag{2}$$

where $p$ is pressure, $p_D$ is the pressure level that the stable layer starts to develop, and $\Delta p$ is its thickness in pressure. Two constants A and B are determined by matching the boundary conditions:

$$\begin{aligned} q(p) &= x\mathrm{NH}_3, \qquad p \to \infty \\ q(p) &= f_x \cdot x\mathrm{NH}_3, \qquad p \leq p_D, \end{aligned} \tag{3}$$

which imply that the ammonia abundance is enriched by a factor of $f_x$ with respect to the deep abundance at pressure levels less than $p_D$. The resulting ammonia profile is displayed in Fig. 4(d), where a small kink in the ammonia profile is visible.

We first examine qualitatively the sensitivity of limb-darkening to these parameters by varying each parameter while holding the others constant. The results are in Fig. 3. The objective is to identify which parameter is constrained by which channel. Our reference state, based on fitting the nadir-only MWR measurements, is $x\mathrm{NH}_3 = 2.7$ times solar, $x\mathrm{H}_2\mathrm{O} = 2.0$ times solar, $\Delta p = 1$ bar, $p_D = 2$ bars, $f_x = 1.2$ and potential temperature $\Theta = 132$ K (referenced to 0.5 bar)[26]. From this reference state, $x\mathrm{NH}_3$ is varied from 2.5 to 2.8 times solar, equivalent to 324 to 376 ppm (Fig. 3a); $x\mathrm{H}_2\mathrm{O}$ is varied from 1.0 to 4.0 times solar (Fig. 3b); $\Delta p$ is varied from 0.25 to 4.0 bars (Fig. 3c); $p_D$ is varied from 1 to 4 bars (Fig. 3d); $f_x$ is varied from 1.0 to 1.4 (Fig. 3e); and $\Theta$ is varied from 130.1 K to 134.1 K (Fig. 3f). Three spectra of the limb-darkening, evaluated at 45 degrees, 30 degrees, and 15 degrees, are plotted against frequency in GHz. Since the width of the shaded area is proportional to the sensitivity of the parameter studied, Fig. 3(c), (d), and (f) show that the limb-darkening is not sensitive to the choices of $p_D$, $\Delta p$, and $\Theta$. Therefore, the precise location, thickness and functional form of the ammonia-variable layer do not affect our retrieved deep ammonia abundance. For example, one may use a linear interpolation for the ammonia profile that starts at pressure level $p_D$ and ends at pressure level $(p_D - \Delta p)$; this does not change our result. Fig. 3(a) shows that the 1.25 GHz channel is most sensitive to the deep ammonia abundance. As demonstrated in Fig. 3(b) and in Janssen et al. (2017)[19], the 2.6 GHz channel exhibits the most sensitivity to the water abundance because the limb-darkening value is largest when the water abundance changes significantly, which is at the water condensation level. Since the limb-darkening of this channel is also affected by $f_x$ (Fig. 3e), there is a potential correlation between the estimation of the water abundance and the ammonia enrichment factor, a reason why $f_x$ must be included in the forward model to allow a conservative estimate of the water abundance. But beecause the 10 GHz channel is more sensitive to $f_x$ than the other channels, it can be used to determine $f_x$ while the 2.6 GHz channel determines $x\mathrm{H_2O}$ and the 1.25 GHz channel determines $x\mathrm{NH_3}$.

We then use a Markov Chain Monte Carlo sampler to assess the posterior probability distribution of the parameters based on the average of all perijoves in the current study observed at 1.25 GHz (24 cm), 2.6 GHz (11.5 cm), 5.2 GHz (5.8 cm), 10 GHz (3 cm), and 22 GHz (1.4 cm) channels (see Methods). Brightness temperatures from the 0.6 GHz channel (50 cm) are not part of this analysis because the opacity of ammonia and water are not known well enough at the greater pressure and temperatures probed with this frequency[27]. In Supplementary Material A we provide a simulation of results for the EZ using the 0.6 GHz channel assuming its opacity is well known, and we obtain a substantial improvement to our precision. The simulation shows that further laboratory data would be useful to constrain the water abundance beyond 30 bars. All mixing ratios in this article are referred to molar mixing ratios and the Solar photospheric abundances are according to Table 1 of Asplund et al. (2009)[28], adjusted to protosolar values[29]. They are also summarized in Supplementary Table 2.

A sample of thermal and compositional profiles explored by the sampler is displayed in Fig. 4, with the median profiles marked in black. The final statistics are expressed by the symbol $\mathrm{A}^{+\sigma}_{-\delta}$, where $\mathrm{A}-\delta, \mathrm{A}, \mathrm{and}\ \mathrm{A}+\sigma$ represent the 16th, 50th and 84th percentiles of the samples (one-sigma uncertainty) in the marginal distribution. The values are relevant for pressures less than about 30 bars where the contribution function of the 1.2 GHz channel peaks. Fig. 5 shows the marginal probability density functions of the ammonia abundance, the water abundance and the ammonia enrichment factor. The distribution of $p_D$ and $\Delta p$ $\mathrm{(not\ shown)}$ are flat, as expected from the sensitivity study; Θ is distributed similarly to the prior, implying that the observation does not contain more information about Θ than the prior (not shown).

The overall estimate of the ammonia abundance is $351^{+22}_{-21}$ ppm, or $2.76^{+0.17}_{-0.16}$ times solar, which is consistent within errors with our previous result, $362^{+33}_{-33}$ ppm[20]. Although the location of this estimate is not directly comparabe to the location of the Galileo measurement, our estimate is at the lower limit of the Galileo Probe Mass Spectrometer's (GPMS) range ($566 \pm 216$ ppm)[10] and is about a factor of two smaller than that suggested by the radio attenuation of the signal from the Galileo Probe ($700 \pm 100$ ppm)[30]. We speculate that the discrepancy may be an overestimation of the radio attenuation due to signal loss by scintillation of the coherent signal caused by atmospheric turbulence. Note that the values used in the ground-based VLA observations (570 ppm and 400 ppm)[24,31] cannot provide an independent estimate of the deep ammonia abundance due to confusion from synchrotron radiation in the foreground at long wavelengths. Rather, the Galileo values are used as the *prior* lower boundary condition and the ammonia abundance is adjusted within the pressure range that the VLA data can be confidently interpreted. Our analysis utilizes the limb-darkening observations unique to the Juno mission and is an original estimation independent of the previous ones.

The overall water abundance is $2.5^{+2.2}_{-1.6} \times 10^3$ ppm, or $2.7^{+2.4}_{-1.7}$ times solar with a long tail toward the larger values. The estimated ammonia and water abundance are negatively correlated because increasing either the ammonia abundance or the water abundance will reduce both the nadir brightness temperature and the limb-darkening (Fig. 6a). Ranges of water abundances as a function of ammonia abundance are summarized in Table 1. Although the centroid of the abundance values for water and ammonia expressed in terms of solar abundance lie on top of each other, we caution that both the magnitude of the error bars and their symmetry differ for the two determinations. In the case of ammonia, the error bars are small and symmetric; in the case of water, our error bars are larger with a tail toward higher values. The potential range of the enrichment ratios for water versus ammonia allows for water to be less enriched than ammonia (solar vs 2.6 times solar) or much more enriched (5 times solar vs 2.9 times solar). Moreover, the quoted uncertainty range encompasses the 16th to 84th percentile of the probability space. A two-sigma (2.5th to 97.5th percentile) uncertainty of the water abundance is $2.7^{+4.8}_{-2.6}$ times solar. Until we are able to utilize the 0.6 GHz channel data, we cannot definitively rule out zero water from fitting the MWR spectra. However, a lower limit of the deep water abundance can be inferred by other evidence such as the existence of lightning[21] or infrared observations. For example, Bjoraker et al. (2018) measured an unexplained cloud opacity at 5 bars in the Great Red Spot, which may indicate a minimum water abundance of $1.1 \times$ solar at shallow depths[32]. In addition, multiple authors[18,33,34] have pointed out that beyond a water molar mixing ratio threshold of about $9 \times 10^3$ ppm, there can be an inhibition of convection and a steep temperature increase at the water condensation level. We have confirmed that the convective inhibition condition does not apply on Jupiter and the moist adiabatic assumption is valid.

Our model successfully reproduces the average MWR measurements of PJ1 to PJ9, both in brightness temperature and limb-darkening. The spatial and temporal variability of different perijoves are treated as the ensemble uncertainty shown in Fig. 7 because we have not accounted for the atmospheric variation. As further discussed in Supplementary Material B, our best-fit model is within one standard deviation of all these constraints. Our solutions for all measurements between latitude 0 and 4°N show no correlation between the residual brightness temperature and the emission angle, implying that there is no obvious missing physical ingredient needed. The ammonia enrichment factor ($f_x$) is approximately distributed between 1.0 and 1.2 with a sharp cut off for values greater than 1.4. Although the inversion result cannot statistically rule out the possibility of $f_x = 1$ (in terms of 2σ errors), including this parameter provides a conservative estimate of the uncertainty of the inferred water abundance. The degeneracy between $f_x$ and the water abundance is shown in Fig. 6(b). $f_x$ is positively correlated with the water abundance because a larger $f_x$ leads to more limb darkening at the 2.6 GHz channel and so does a lesser water abundance. Since no previous studies have indicated the existence of such a region, the estimation of the water abundance is $1.7^{+1.6}_{-1.2}$ times solar given $f_x = 1$, and the corresponding probability density function is shown in Fig. 6(c).

Historically, there has been a strong interest in determining Jupiter's global elemental abundances to aid in our understanding of Jupiter and solar system formation theories. If the value in the EZ is the global value, then (1) at the one-sigma level our result is inconsistent with any model of the planetesimal enrichment of Jupiter's envelope that postulates the Galileo probe depletion of O relative to solar to be a global depletion. For example, Lodders et al. constructed a model in which Jupiter's envelope is fed by planetesimals greatly depleted in water and enriched in carbon, so that the total oxygen abundance is only 0.47 times solar[35]. (2) Conversely, planetesimal models in which the trapping of volatiles in a clathrate hydrate from a solar composition disk produces water enrichments 10 times solar[14], supported by observation of the atmospheric waves generated by the Shoemaker-Levy 9 impacts[36], are ruled out. However, a number of other published models produce water enrichments consistent with those derived here, including the trapping of volatiles in cold amorphous ice leading to an enrichment of O consistent with that of the other major elements and heavy noble gases[12,13] and roughly three times solar O abundance yielded by the equatorial wave analysis[37]. Because our water abundance pertains to the equator only, we shall postpone a thorough discussion of the previous models to a dedicated article in the future.

We have shown that the structure of Jupiter's Equatorial Zone is steady, relatively uniform vertically and close to a moist adiabat, based on which, we derived its water abundance. The thermal structure outside of the equator is still ambiguous due to the non-uniform distribution of ammonia gas, of which we do not know its physical origin. Deriving the thermal structure outside of the equator in the future not only hints about the water abundance on Jupiter at other latitudes but also places constraints on the atmospheric circulation model for giant planets in the solar system and beyond.

Table 1. The first three rows show the conditional distribution of the water abundance given the ammonia abundance. The last row shows the overall distribution of ammonia and water. The protosolar values are according to Atreya et al. (2019)[29] and summarized in Supplementary Table S1. These results are based on 1.25 to 22 GHz data from Juno MWR probing approximately 0.7 to 30 bars pressure.

| Ammonia (protosolar) | Ammonia (ppm) | Water (protosolar) | Water (1000 ppm) |
|---|---|---|---|
| **2.6** | 330 | $4.1^{+2.1}_{-2.1}$ | $3.8^{+1.9}_{-1.9}$ |
| **2.8** | 356 | $2.4^{+2.0}_{-1.5}$ | $2.2^{+1.9}_{-1.4}$ |
| **3.0** | 381 | $1.6^{+1.4}_{-1.0}$ | $1.5^{+1.3}_{-0.9}$ |
| $2.76^{+0.17}_{-0.16}$ | $351^{+22}_{-21}$ | $2.7^{+2.4}_{-1.7}$ | $2.5^{+2.2}_{-1.6}$ |

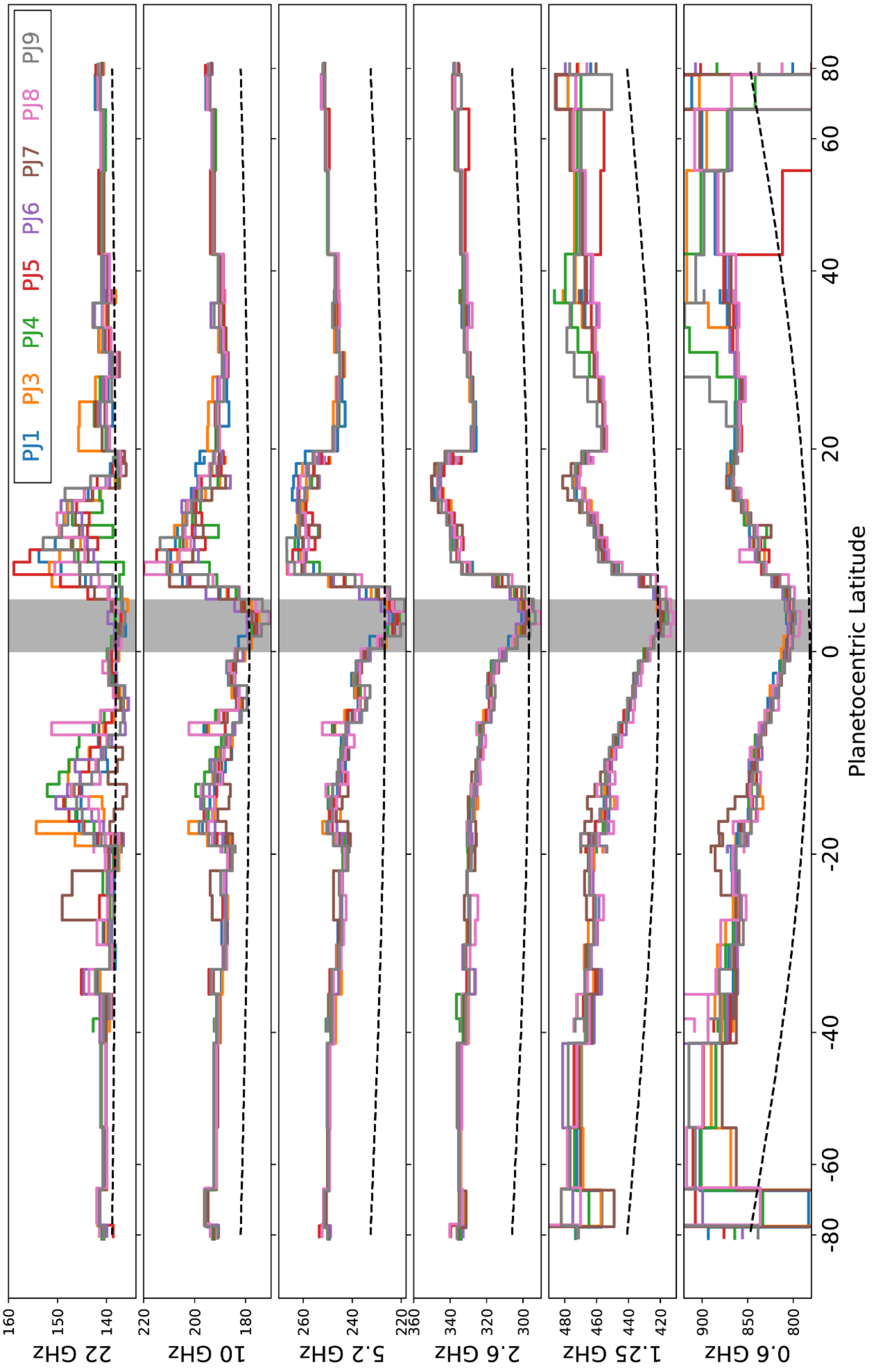

Figure 1. Nadir brightness temperatures of perijoves PJ1 to PJ9. The abscissa is planetocentric latitude, scaled by the sine function. Brightness temperatures (the ordinate) of MWR channels at 0.6 GHz – 22 GHz are ordered from bottom to top. The perijoves are distinguished by different colors indicated in the figure. Small step functions in each curve represent the resolution element in latitude, which is hand-tuned using PJ1 data (about 1 degree from 20$^o$S to 20$^o$N, 2 degrees from 20$^o$ to 40$^o$ north and south, and 8 degrees from 40$^o$ to 90$^o$ north and south). Black dashed lines are reference nadir brightness temperatures of an ideal moist adiabatic profile with potential temperature being 132 K referenced at 0.5 bar, and with ammonia and water abundance being 2.7 times solar.

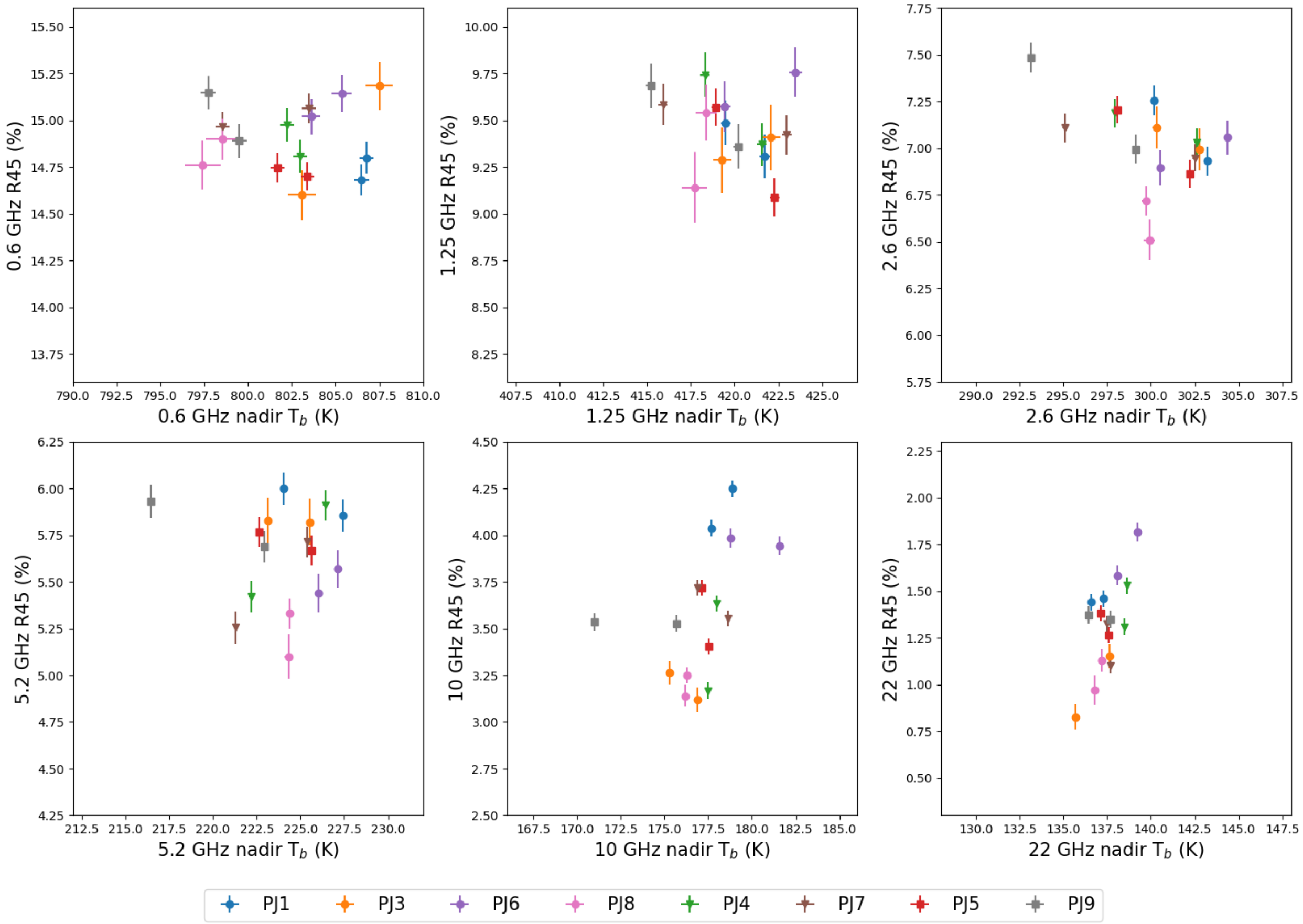

Figure 2. Limb-darkening versus nadir brightness temperature for each channel in the North Equatorial Zone (0°-4°N). The vertical axis of each panel spans 2% in limb-darkening evaluated at 45° emission angle. Each perijove has two data points. One is averaged between 0-2°N and the other is averaged between 2-4°N. The color for each perijove is the same as that in Figure 1. PJ1, 3, 6, and 8 are gravity orbits, indicated by circles. PJ4 and 7 are MWR orbits, indicated by triangles. PJ5 and 9 are MWR tilt orbits, indicated by squares. The error bar of each data point represents the estimated error taking into account instrument noise and the deconvolution of antenna beams.

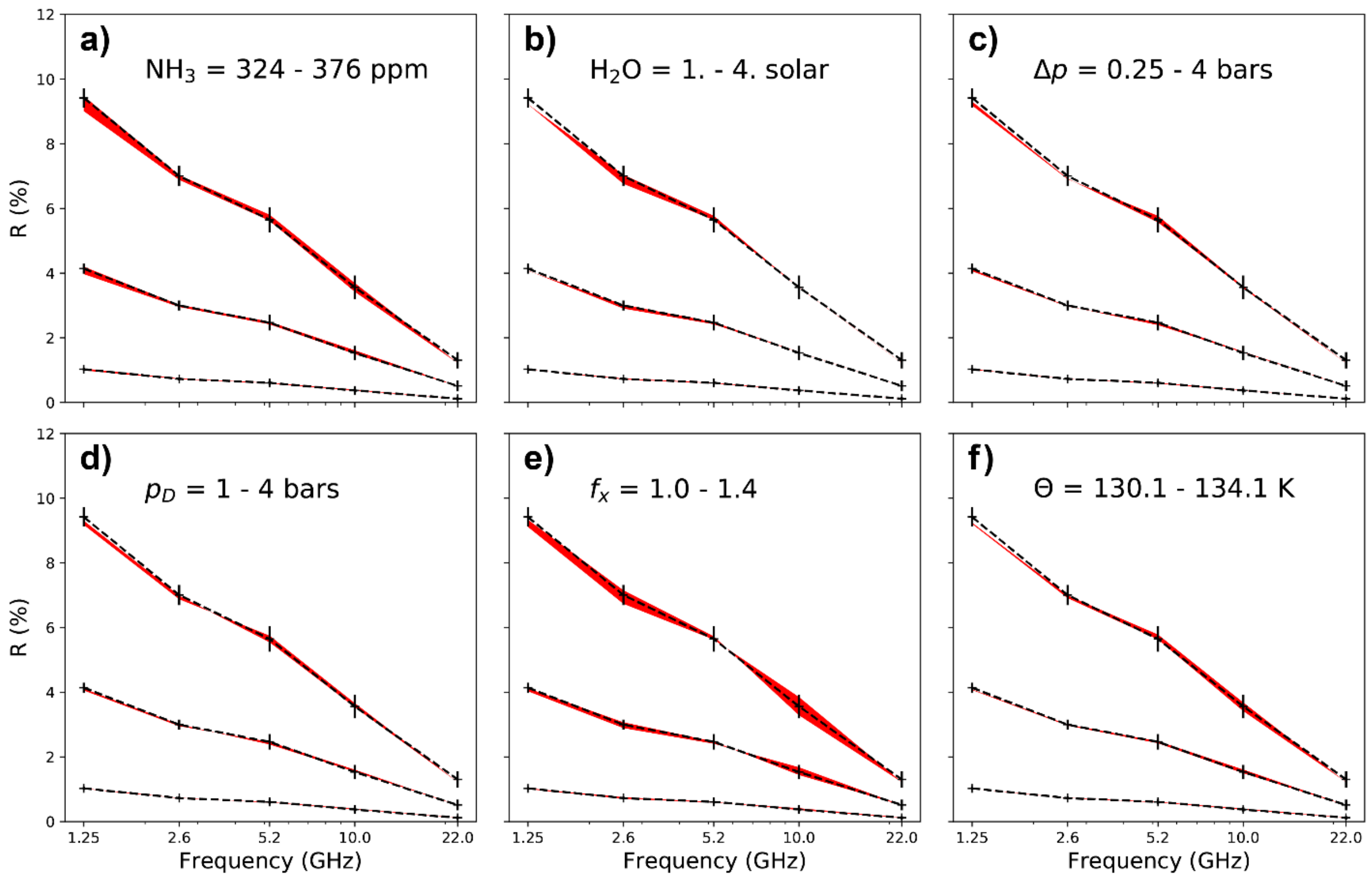

Figure 3. Sensitivity study of model parameters. The reference profile is constructed by the state vector *X=(350 ppm, 2.0* solar*, 1* bar*,2* bars*,1.2,132.1* K*)*. Three black lines in each panel show the limb-darkening at 45°, 30° and 15° emission angles respectively from top to bottom. The red shaded area shows the range of the limb-darkening when each parameter is varied within the values indicated in the figure.

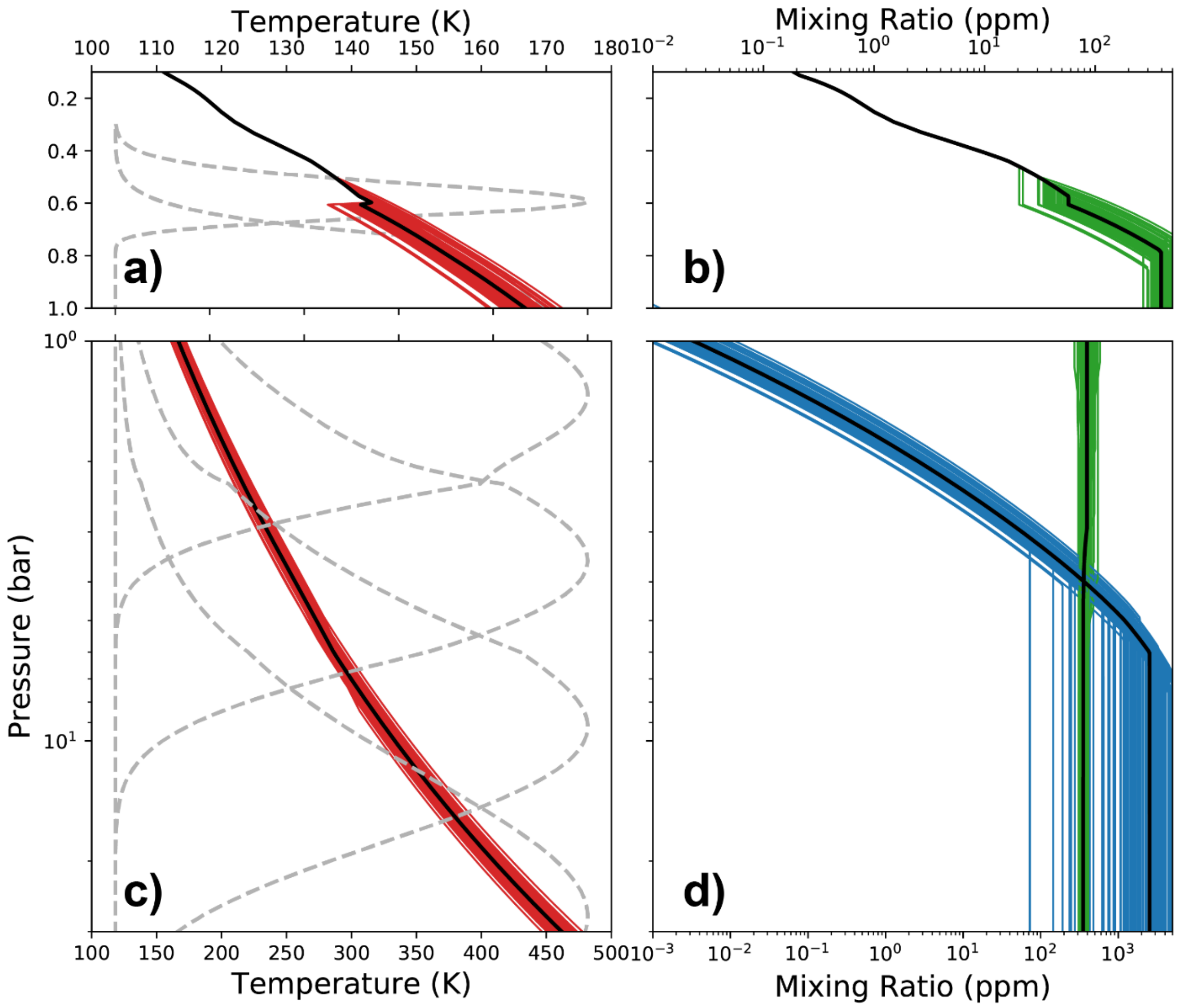

Figure 4. a) Temperature profile above 1 bar pressure level and normalized contribution functions of 22 GHz channel and 10 GHz channel. b) Ammonia profile above 1 bar pressure level. c) Temperature profile from 30 bars to 1 bar and normalized contribution functions of 10 GHz, 5.2 GHz, 2.6 GHz, 1.2 GHz from top to bottom. d) Ammonia and water profile from 30 bars to 1 bar. The black lines indicate the median atmospheric model in all samples. The colored lines (red for temperature, blue for water and green for ammonia) show 100 random samples drawn from the whole probability distribution.The profiles above ~0.6 bar are replaced with the CIRS obsrevations[38], which sometimes leads to a discontinuity at this pressure level.

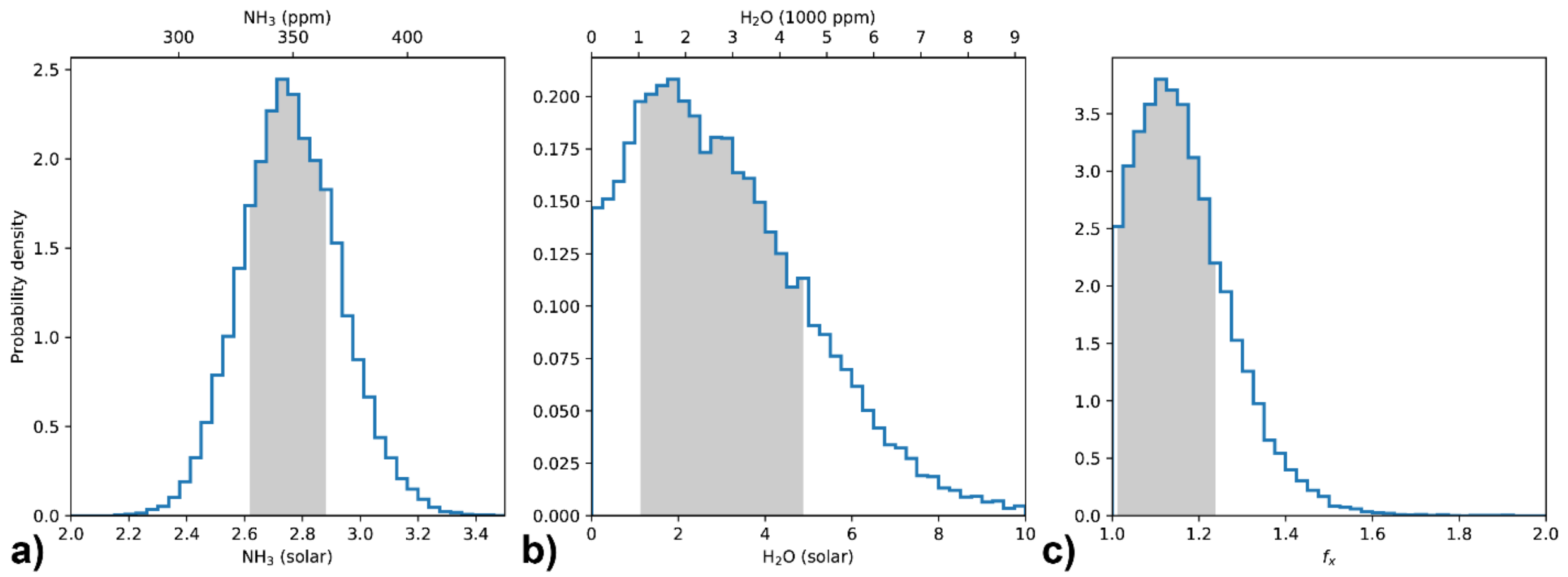

Figure 5. Probability density functions of a) ammonia, b) water, and c) ammonia enrichment factor estimated by a Markov Chain Monte Carlo sampler using the MWR data from latitudes 0-4 degrees north. The shaded area marks the $1\sigma$ confidence interval (16th to 84th percentile). A study using synthetic data is provided in Supplementary Fig. A2 for comparison.

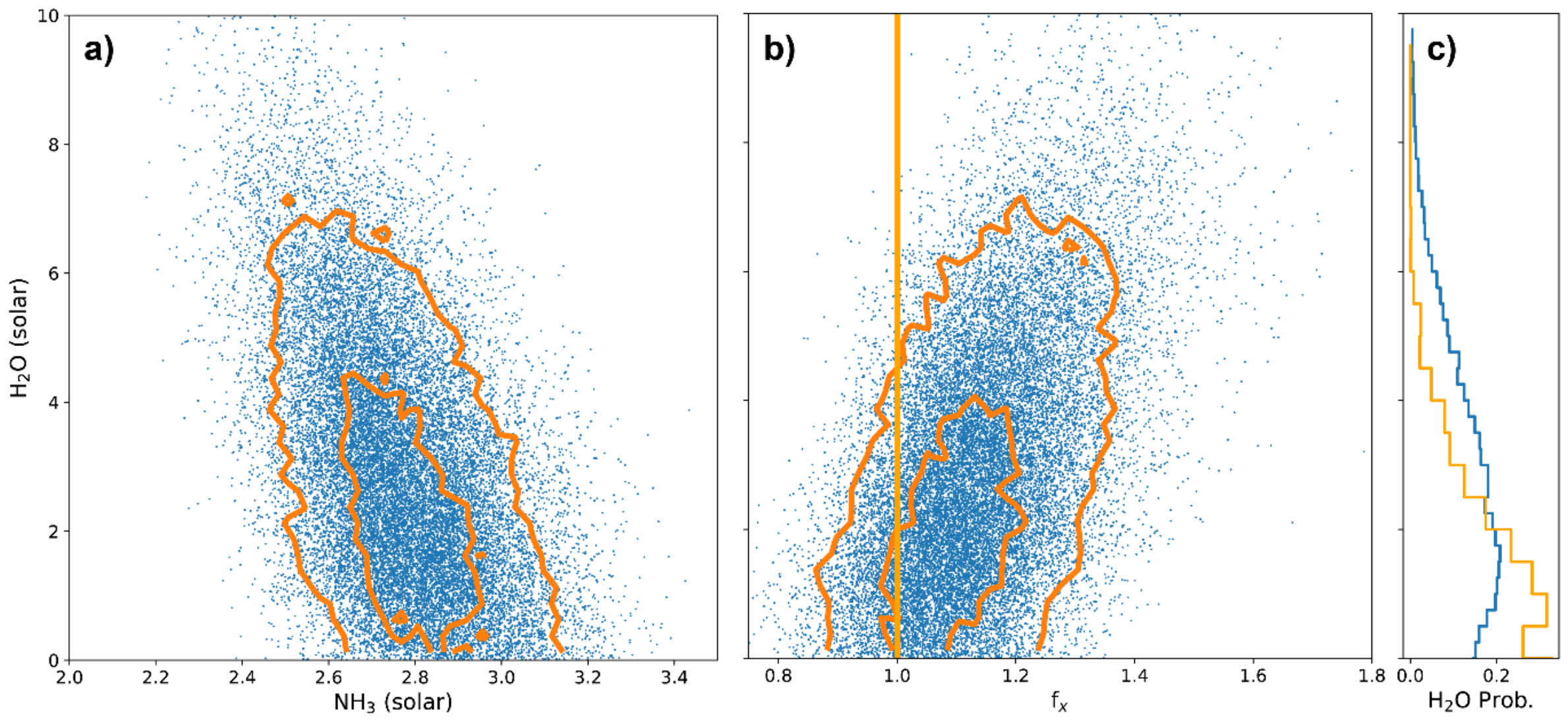

Figure 6. Correlation between a) the water abundance and the ammonia abundance and b) the ammonia enrichment factor ($f_x$) and the water abundance. Each dot represents one state in the Markov Chain. The density of the dots is proportional to the probability. Two contours show $1\sigma$ and $2\sigma$ confidence intervals. Panel c) shows the overall marginal probability distribution for the water abundance (blue line) and the marginal probability distribution for the water abundance assuming $f_x = 1$ (orange line). The blue line is the same as that in Fig. 5 (b).

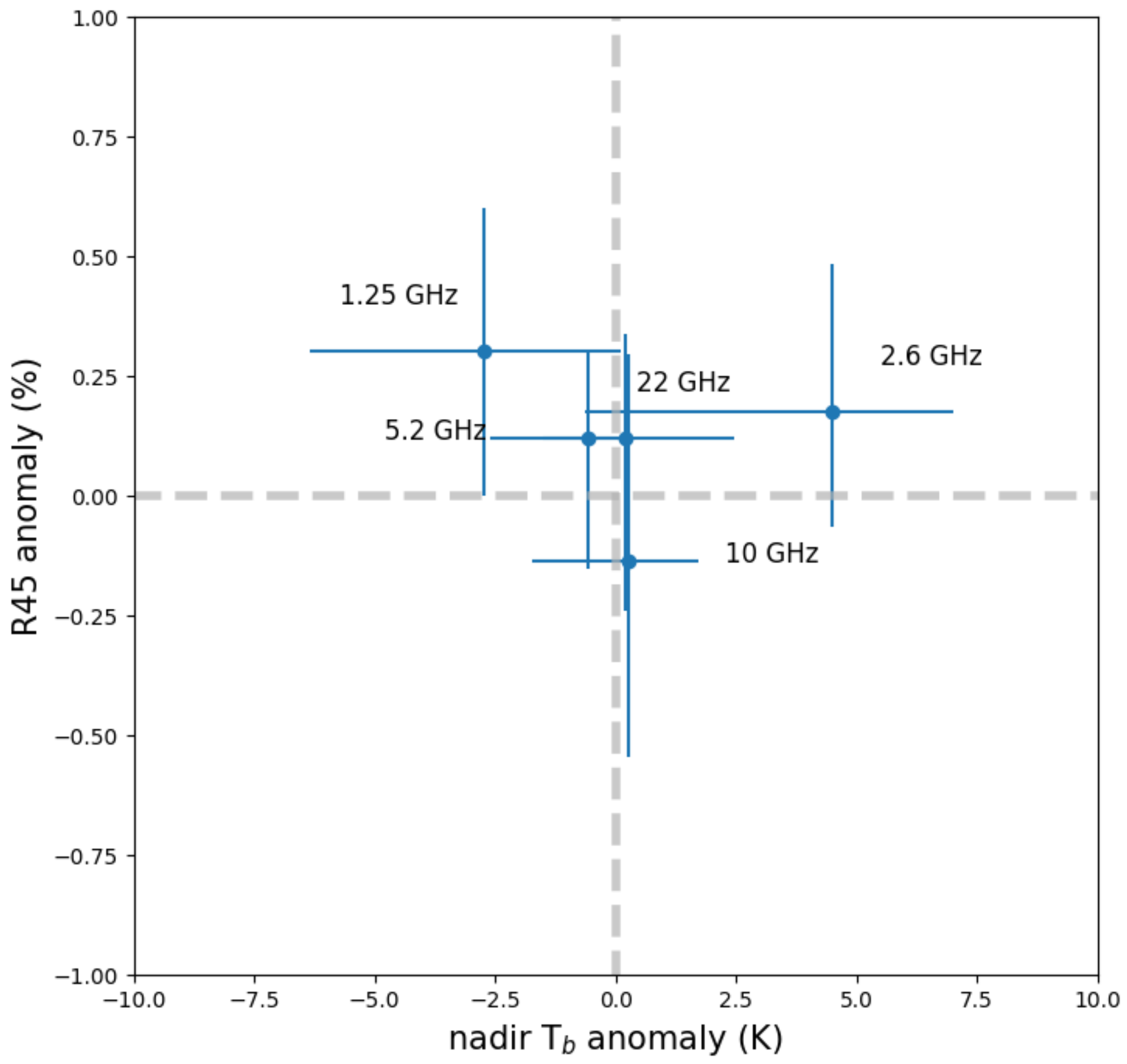

Figure 7. Measured brightness temperature and limb-darkening anomalies for 1.25 – 22 GHz channels with respect to the best-fitted model obtained from Juno measurements in the North Equatorial Zone (0°-5°N). The error bars represent overall uncertainty estimates including atmospheric variability, deconvolution, instrument noise, and instrument calibration.

## Methods.

### 1. Deconvolution methods

The brightness temperature at microwave channel $\lambda$, latitude $\phi$, and emission angle $\mu = \cos(\theta)$ is parameterized by a quadratic representation of the form

$$T_\lambda(\phi,\mu) = \xi_\lambda(\mu)(a_\lambda(\phi) + b_\lambda(\phi)(1-\mu) + c_\lambda(\phi)(1-\mu)^2), \quad \text{(M1)}$$

where $\xi_\lambda(\mu)$, different for each channel λ, is given by the ratio of the brightness temperature of a model atmosphere and a fit to that model atmosphere only over emission angles less than 53 degrees ($\mu < 0.6$). This ratio equals to one for angles less than 53 degrees but deviates from unity at larger angles. The idea is to fit small angles with a quadratic function but to assume some profile for large angles, which contribute only a small fraction to the antenna temperatures included in the measurement set. For this work, only emission angles less than 45 degrees are used. Therefore, the prefactor $\xi_\lambda(\mu)$ is omitted in the main manuscript.

The performance of the fit was evaluated using an equilibrium cloud condensation model[1], and it was found that the fits were not strongly dependent on the specific model atmosphere chosen to determine $\xi_\lambda(\mu)$. It was found that a quadratic fit of the form given above ($T_B^{(fit)}$) was sufficient to fit the limb-darkening at 45 degrees (evaluated using a real radiative transfer program, $T_B^{(model)}$) for 1.25 – 22 GHz channels to within an absolute error of 0.1% (Extended Fig. E1) and the nadir brightness temperature to within a relative value of 0.1% (Extended Fig. E2) over all perijoves up to PJ9 and over a range of model atmospheres assuming both dry and moist adiabats for all latitudes and spanning ammonia abundances of 2.1 to 3.3 times solar, water abundances of 0.5 to 8 times solar, and half-bar reference temperatures of 130 K to 135.6 K. The maximum errors for 0.6 GHz channel, not included in this analysis, were about twice as large. More details of the deconvolution algorithm, including the treatment of the synchrotron radiation will be summarized in a separate paper.

The measured antenna temperatures are a convolution of the antenna pattern for each channel with brightness temperatures that depend on both emission angle and spatial location. To determine the brightness temperatures a deconvolution must be performed to negate the blurring induced by the instrument antenna pattern. Because the convolution is a linear operation, the set of measured antenna temperatures, $T_a$, can be represented as the product of a linear operator $M$ with a set of coefficients, $X$, where the components of $X$ represent information regarding the angular dependence of the brightness temperature at different locations on Jupiter (the $a_\lambda(\phi), b_\lambda(\phi), c_\lambda(\phi)$ coefficients in equation (M1). Then we want to find the optimal solution $X^*$ that minimize the cost function:

$$(T_a - MX)^T S_b^{-1}(T_a - MX) + (X - X_p)S_a^{-1}(X - X_p)\,, \tag{M2}$$

which is:

$$X^* = X_p + (M^T S_b^{-1} M + S_a^{-1})^{-1}\left(M^T S_b^{-1}\left(T_a - MX_p\right)\right), \tag{M3}$$

where $X_p$ is an assumed prior, $S_b$ is the measurement covariance matrix, $S_a$ is a prior covariance, and $T_a$ is the set of measured antenna temperatures, modified to estimate the contribution from synchrotron radiation. The covariance of $X^*$ is a sparse matrix whose sparsity pattern is shown in Extend Fig. E3 for the 10 GHz channel. The other channels are similar. The deconvolution process is performed in two steps. First, a coarse grid of 1.2 degree spacing over the latitude range presented in this work for a and 2.4 degrees for b and c is used without any prior estimate $X_p$. This step is often sufficient to yield residuals that are within a small multiple of the instrument noise level. The coarse grid solution can then be used as prior estimate for a refined solution on a finer grid to reduce the $\chi^2$ to a value close to 1.

### 2. Spectral inversion methods

Let the parameter vector $X = (x\mathrm{NH_3}, x\mathrm{H_2O}, f_x, \Delta p, p_D, \Theta)^T$ and the measurement vector $Y = (a_1, b_1, c_1, a_2, b_2, c_2 \ldots, a_5, b_5, c_5)^T$, where $a, b$ and $c$ are brightness temperature coefficients introduced in equation (M1) and their subscripts are channel numbers. We accounted for the limb-darkening via the use of these coefficients. The sensitivity study in the previous section has demonstrated that the 2.6 GHz channel best constrains the water abundance, which is the main objective of this article. We use a Markov Chain Monte Carlo (MCMC) sampler[2] to assess the joint probability of the parameter $X$ given the observation $Y$. Bayes' theorem states that:

$$p(X|Y) = \frac{P(Y|X)P(X)}{P(Y)}, \tag{M4}$$

where $P(X)$ is the prior distribution of the parameter, $P(Y|X)$ is the probability of observing Y given $X$, and $p(X|Y)$ is the probability of the parameter $X$ given the observation $Y$. We assume the prior probability of Θ is a Gaussian distribution with a mean of 132.1 K and a standard deviation of 2 K[3]. The prior probabilities of other parameters are uniformly distributed. Denoting the forward model as $F(X)$ and assuming Gaussian statistics, $P(Y|X)$ is:

$$p(Y|X) = \frac{1}{\sqrt{(2\pi)^d|\Sigma|}} \exp\left(-\frac{1}{2}\big(Y - F(X)\big)^T \Sigma^{-1}(Y - F(X))\right), \tag{M5}$$

where $d = 15$ is the dimension of the observation vector $Y$, and $\Sigma$ is the formal covariance matrix estimated from the deconvolution process. $\Sigma$ is a block diagonal matrix with each block of size $3 \times 3$ because the five channels are independent of each other. The magnitude of the elements in $\Sigma$ is illustrated by the error bars in Fig. 7. The MCMC sampling algorithm uses 24 Markov chains with 6,000 states in each chain to fit the measurement vector.

The thermal profile is obtained by integrating the moist adiabatic temperature gradient upward to approximately 0.6 bar[4]. Then the temperature profile is replaced by the Cassini/CIRS observations in the upper troposphere and stratosphere[5]. Similarly, the compositional profiles are constant up to the condensation levels of each species, except for ammonia, which is modified slightly according to the description in the main text. Then their abundances follow their saturation vapor curves assuming 100% relative humidity and are replaced with the Cassini/CIRS observations for pressures less than 0.6 bar.

## Methods references

**Acknowledgments.** The research was carried out at the Jet Propulsion Laboratory, California Institute of Technology, under a contract with the National Aeronautics and Space Administration. The Juno mission and the team members at JPL were supported by NASA grant NNN12AA01C. TG acknowledges support from CNES. We thank all Juno team members for the collaborative efforts and three anonymous reviewers for insightfull suggestions to improve the manuscript.

**Materials & Correspondence.** We note that Juno MWR data can be accessed on the Planetary Data System (PDS) https://pds.nasa.gov/. All other data or materials that are presented in the paper but not archived in the PDS should be addressed to C.L. (cli@gps.caltech.edu).

**Author Contributions.** C.L. developed the inversion software and performed the data analysis. All authors discussed the results and commented on the manuscript.